\documentclass[aps,pre,showpacs,floats,preprint]{revtex4}
\usepackage{amssymb}
\usepackage{graphics}
\usepackage[dvipdfm]{graphicx}
\usepackage{epsfig}
\usepackage{dcolumn}
\usepackage{amsmath}

\begin{document}

\title{Maxwell Times in Higher-Order Generalized Hydrodynamics: Classical Fluids, and Carriers and Phonons in
Semiconductors}
\author{Cl\'{o}ves G. Rodrigues$^{1}$, Carlos A. B. Silva$^{2}$,
Jos\'{e} G. Ramos$^{3}$, Roberto Luzzi$^{3}$\footnote{group home
page: www.ifi.unicamp.br/aurea; email: cloves@pucgoias.edu.br}}
\affiliation{ $^{1}$Departamento de F\'{\i}sica, Pontif\'{\i}cia
Universidade Cat\'{o}lica de Goi\'{a}s, 74605-010 Goi\^{a}nia,
Goi\'{a}s, Brazil\\
$^{2}$Instituto Tecnol\'{o}gico de Aeron\'{a}utica, 12228-901
S\~{a}o Jos\'{e} dos Campos, SP, Brazil\\
$^{3}$Condensed Matter Physics Department, Institute of Physics
``Gleb Wataghin'' State University of Campinas-Unicamp, 13083-859
Campinas, SP, Brazil}

\date{\today }

\begin{abstract}
A family of the so-called Maxwell times which arises in the context
of Higher-Order Generalized Hydrodynamics (also called Mesoscopic
Hydro-Thermodynamics) is evidenced. This is done in the framework of
a HOGH build within a statistical foundation in terms of a
Non-Equilibrium Statistical Ensemble Formalism. It consists in a
description in terms of the densities of particles and energy and
their fluxes of all orders, with the motion described by a set of
coupled nonlinear integro-differential equations involving them.
These Maxwell Times have a fundamental role in determining the type
of hydrodynamic motion that the system would display in the given
condition and constraints. The different types of motion are well
described by contractions of the full description done in terms of a
reduced number of fluxes up to a certain order.
\end{abstract}


\maketitle


\section{Introduction}

We present a generalization of Maxwell time [1], which can also be
referred-to as generalized Maxwellian relaxation time [2], a family
of which is present in the evolution equations of Higher-Order
Hydrodynamics, also dubbed as Mesoscopic Hydro-Thermodynamics (MHT)
[3-6]. Such generalized Maxwell times are associated to the
dampening of densities of particles (molecules in a classical fluid)
or quasi-particles (Bloch-band electrons, harmonic phonons, magnons,
polaritons, etc., in solid state matter, in semiconductors, for
example), and of their energy density, together with the fluxes of
all orders of both.

The origin of Maxwell times goes back to the fundamental article by
J. C. Maxwell in 1867 on the dynamical theory of gases [1], in what
is related to viscoelasticity and which is being expressed as the
quotient of the dynamical viscosity coefficient with the modulus of
rigidity. According to Maxwell it may be called the "time of
relaxation" of the elastic force. It has been noticed [2] that given
a fluid subjected to some variable external forces, which vary
periodically in time with frequency $\omega$, of the period
$1/\omega$ is large compared with Maxwell time , i.e. $\omega \theta
\ll 1$, the fluid under consideration will be have as an ordinary
viscous fluid. If however, the frequency $\omega$ is sufficiently
large (so that $\omega \theta \gg 1$, the fluid will be have as an
amorphous solid.

We present a detailed study of the generalized Maxwell times that
arise in a Mesoscopic Hydro-Thermodynamics, built upon a
Non-Equilibrium Statistical Ensemble Formalism (NESEF for short)
[7-10], at the classical and quantum mechanical levels of HOGH
described in Refs. [5,6,11,12].

In Section II is considered the classical MHT of a fluid of
particles embedded in a thermal bath. In Section III the case of a
MHT at the quantum level of a system of phonons in semiconductors,
and in Section IV a MHT at the quantum level of a system of
electrons in doped semiconductors.

\newpage
\section{Maxwell Times in the MHT of a Classical Fluid}

Construction of MHT in the framework of a Nonequilibrium Statistical
Ensemble Formalism is presented in Ref. [5]. There it is provided an
approach enabling for the coupling and simultaneous treatment of the
kinetic and hydrodynamic levels of descriptions. It is based on a
complete thermo-statistical approach in terms of the densities of
matter and energy and their fluxes of all orders, as well as on
their direct and cross correlations, covering systems arbitrarily
far-removed from equilibrium. The set of coupled nonlinear
integro-differential hydrodynamic equations is derived. They are the
evolution equations of a Grad-type approach involving the moments of
all orders of the single-particle distribution
$f_{1}(\mathbf{r},\mathbf{p};t)$, derived from a generalized kinetic
equation, built in the framework of a Nonequilibrium Statistical
Ensemble Formalism [13].

The moments of the single-particle distribution function, in
momentum space $\mathbf{p}$, are the hydrodynamic variables
%
\begin{equation}
n(\mathbf{r},t) = \int d^{3}pf_{1}(\mathbf{r},\mathbf{p};t) \, ,
\end{equation}
which is the densities of particles,
%
\begin{equation}
\mathbf{I}_{n}(\mathbf{r},t) = \int d^{3}p \,
\mathbf{u}(\mathbf{p})f_{1}(\mathbf{r},\mathbf{p};t) \, ,
\end{equation}
with
%
\begin{equation}
\mathbf{u}(\mathbf{p}) = \frac{\mathbf{p}}{m} \, ,
\end{equation}
where $\mathbf{I}_{n}$ is the flux (current) of particles,
%
\begin{equation}
I_{n}^{[2]}(\mathbf{r},t) = \int d^{3}p \,
\mathbf{u}^{[2]}(\mathbf{p}) f_{1}(\mathbf{r},\mathbf{p};t) \, ,
\end{equation}
where $\mathbf{u}^{[2]} = [\mathbf{u} \, \colon \mathbf{u}]$ is the
inner tensorial product of vectors $\mathbf{u}$, with $I_{n}^{[2]}$
being the second-order flux (or flux of the flux), a rank-2 tensor,
which multiplied by the mass is related to the pressure tensor, and
%
\begin{equation}
I_{n}^{[\ell]}(\mathbf{r},t) = \int d^{3}p \,
\mathbf{u}^{[\ell]}(\mathbf{p}) f_{1}(\mathbf{r},\mathbf{p};t) \, ,
\end{equation}
are the higher-order fluxes of order $\ell \geqslant 3$ (the
previous three of Eqs. (1), (2) and (4), are those for $\ell = 0,1$
and $2$ respectively) where $\mathbf{u}^{[\ell]}$ is the $\ell$-rank
tensor consisting of the inner tensorial product of $\ell$ vectors
$\mathbf{u}$ of Eq. (3), that is,
%
\begin{equation}
\mathbf{u}^{[\ell ]}(\mathbf{p}) = \left[ \frac{\mathbf{p}}{m}
\colon \frac{\mathbf{p}}{m}\ldots (\ell \mathrm{-times}) \ldots
\colon \frac{\mathbf{p}}{m}\right] \, .
\end{equation}

We do have what can be called the \emph{family of hydrodynamical
variables describing the material motion}, i.e., the set
%
\begin{equation}
\{n(\mathbf{r},t); \, \mathbf{I}_{n}(\mathbf{r},t); \,
\{I_{n}^{[\ell ]}(\mathbf{r},t)\} \} \, ,
\end{equation}
with $\ell = 2, 3, \ldots$, which we call the MHT-family $n$.

On the other hand, we do have the \emph{family of hydrodynamical
variables describing the thermal motion}, which we call the
MHT-family $h$, consisting of
%
\begin{equation}
h(\mathbf{r},t) = \int d^{3}p \, \frac{p^{2}}{2m}
f_{1}(\mathbf{r},\mathbf{p};t) \, ,
\end{equation}
%
\begin{equation}
\mathbf{I}_{h}(\mathbf{r},t) = \int d^{3}p \, \frac{p^{2}}{2m}
\frac{\mathbf{p}}{m} f_{1}(\mathbf{r},\mathbf{p};t) \, ,
\end{equation}
%
\begin{equation}
I_{h}^{[\ell ]}(\mathbf{r},t) = \int d^{3}p \, \frac{p^{2}}{2m}
\mathbf{u}^{[\ell ]}(\mathbf{p}) f_{1}(\mathbf{r},\mathbf{p};t) \, ,
\end{equation}
with $\ell = 2, 3, \ldots$, that is, in compact form, those in the
set
%
\begin{equation}
\{h(\mathbf{r},t); \, \mathbf{I}_{h}(\mathbf{r},t); \,
\{I_{h}^{[\ell ]}(\mathbf{r},t)\}\} \, ,
\end{equation}
which are, respectively, the density of energy, its first vectorial
flux (heat current), and the higher-order tensorial fluxes. It can
be noticed that in this case of a parabolic type energy momentum
dispersion relation, $E(p)=p^{2}/2m$, the set of Eq. (11) is
encompassed in the previous one: in fact
%
\begin{equation}
h(\mathbf{r},t) = \frac{m}{2}Tr\{I_{n}^{[2]}(\mathbf{r},t)\} \, ,
\end{equation}
%
\begin{equation}
\mathbf{I}_{h}(\mathbf{r},t) =
\frac{m}{2}Tr_{2}\{I_{n}^{[3]}(\mathbf{r},t)\} \, ,
\end{equation}
where $Tr_{2}$ stands for the contraction of the first two indexes,
and, in general
%
\begin{equation}
I_{h}^{[\ell ]}(\mathbf{r},t) = \frac{m}{2}Tr_{2}\{I_{n}^{[\ell
+2]}(\mathbf{r},t)\} \, ,
\end{equation}
for all the other higher-order fluxes of energy, that is, any flux
of energy of order $\ell$ is contained in the flux of matter of
order $\ell + 2$.

Let us consider the equations of evolutions for the basic
macrovariables of the family of particle motion, which for the
general flux of order $\ell$ ($\ell =0,1,2,\ldots $) is
%
\begin{equation}
\frac{\partial}{\partial t}I_{n}^{[\ell]}(\mathbf{r},t) = \int
d^{3}p \, \mathbf{u}^{[\ell ]}(\mathbf{p}) \frac{\partial}{\partial
t} f_{1}(\mathbf{r},\mathbf{p};t) \, ,
\end{equation}

Using in Eq. (15) the evolution equations for
$f_{1}(\mathbf{r},\mathbf{p};t) $ of Ref. [13] it follows the
general set of coupled equations for the density, $\ell =0$, the
current, $\ell =1$, and all the other higher-order fluxes, $\ell
\geqslant 2$, given by
%
\begin{equation*}
\frac{\partial }{\partial t}I_{n}^{[\ell ]}(\mathbf{r},t)+\nabla
\cdot I_{n}^{[\ell +1]}(\mathbf{r},t) =
\end{equation*}
%
\begin{eqnarray}
&=&-\frac{1}{m}\sum\limits_{s=1}^{\ell }\sigma
(1,s)[\mathcal{F}(\mathbf{r},t)I_{n}^{[\ell
-1]}(\mathbf{r},t)]-\theta _{n\ell }^{-1}I_{n}^{[\ell ]}(\mathbf{r},t)+  \notag \\
&&a_{L0}\sum\limits_{s=1}^{\ell }\sigma (1,s)[\nabla I_{n}^{[\ell
-1]}(\mathbf{r},t)]+2\ell a_{L1}\nabla \cdot I_{n}^{[\ell
+1]}(\mathbf{r},t)  +  \notag \\
&&J_{NL}^{[\ell ]}(\mathbf{r},t)+S_{n}^{[\ell ]}(\mathbf{r},t) \, .
\end{eqnarray}

The last term on the right of Eq. (16) is given by
%
\begin{eqnarray}
S_{n}^{[\ell ]}(\mathbf{r},t) &=&b_{\tau 0}\{\sigma
(1,s)[1^{[2]}I_{n}^{[\ell -2]}(\mathbf{r},t)\} + b_{\tau
1}\frac{2}{m}\{\sigma
(1,s)[1^{[2]}I_{h}^{[\ell -2]}(\mathbf{r},t)\}  \notag \\
&&+3\ell a_{\tau 1}\frac{2}{m}I_{h}^{[\ell
]}(\mathbf{r},t)+a_{L1}\frac{2}{m}\sum\limits_{s=1}^{\ell }\sigma
(1,s)[\nabla I_{h}^{[\ell -1]}(\mathbf{r},t)]+R_{n}^{[\ell
]}(\mathbf{r},t) \, ,
\end{eqnarray}
with all quantities involved given in Ref. [5]: we omit the details
that are not necessary for the analysis of Maxwell times.

The first contribution on the right of Eq. (17) contain the flux of
particles of order $\ell -2$, whereas the other three contributions
are associated to the fluxes of energy of orders $\ell -2$, $\ell
-1$ and $\ell$, terms that can be considered as associated to
thermo-striction effects which couple these equations with the set
of kinetic equations describing the motion. However, it can be
noticed the already mentioned fact that the fluxes of energy can be
given in terms of those of particles, as given in Eqs. (12) to (14).

The second term on the right of Eq. (16), for the evolution of the
flux of order $\ell$, is the one that contains the Maxwell time
$\theta_{n\ell}$ associated to such flux, namely
%
\begin{equation}
\theta _{n\ell }^{-1} = -\ell |a_{\tau 0}|+\ell (\ell -1)|b_{\tau
1}| \, ,
\end{equation}
with the coefficients $a_{\tau 0}$ and $b_{\tau 1}$ given by
%
\begin{equation}
a_{\tau 0} = \frac{\mathcal{V}}{(2\pi )^{3}} \frac{4\pi }{3} \int dQ
\, Q^{4} f_{\tau 0}(Q) \, ,
\end{equation}
with
%
\begin{equation}
f_{\tau 0}(Q) = - \frac{n_{R}M\beta _{0}^{3/2}\pi
}{\mathcal{V}\sqrt{2\pi }m^{2}} \frac{|\psi (Q)|^{2}}{Q}\left(
\frac{m}{M}+1\right) \, ,
\end{equation}
where $\psi (Q)$ is the Fourier transform of the potential energy
$w(|\mathbf{r}_{j}-\mathbf{R}_{\mu}|)$, between the $j$-th particle
at position $\mathbf{r}$ and the one in the thermal bath at position
$\mathbf{R}_{\mu}$, $n_{R}$ is the density of particles in the
thermal bath, $\mathcal{V}$ is the volume, and $\beta
_{0}^{-1}=k_{B}T_{0}$, with $T_{0}$ being the temperature of the
thermal bath, moreover
%
\begin{equation}
b_{\tau 1} = - \frac{a_{\tau 0}}{5}\left( \frac{m}{M} + 1
\right)^{-1} \, ,
\end{equation}
and $m$ and $M$ are the mass of the particles and of the particles
in the thermal bath respectively. According to Eqs. (18) and (21) we
can write
%
\begin{equation}
\theta _{n\ell }^{-1} = - \ell |a_{\tau 0}|\left[ 1+\frac{1}{5}(\ell
-1)\left( \frac{M}{m+M}\right) \right] \, ,
\end{equation}
and it can be noticed that $|a_{\tau 0}|$ is the reciprocal of the
Maxwell time of the first flux which is the reciprocal of the
relaxation time of the linear momentum $\mathbf{p}(\mathbf{r},t)$,
once $\mathbf{p}(\mathbf{r},t) = m\mathbf{I}_{n}(\mathbf{r},t)$.

Maxwell times of Eq. (18) have their origin from the collision
integral $J_{\tau}^{(2)}(\mathbf{r},\mathbf{p};t)$ in the kinetic
equation for the single-particle distribution present in Eq. (A.11)
in Appendix A in Ref. [13]. It arises out of the interaction with
the thermal bath in the contributions called effective friction
force and diffusion in momentum space.

In a recurrent procedure it follows from Eq. (18) that we can write
%
\begin{equation}
\theta _{n\ell }^{-1} = \ell \left[ 1+ \frac{M}{5(m+M)}(\ell
-1)\right] \theta _{n1}^{-1} \, ,
\end{equation}
what tell us that any characteristic time for $\ell \geqslant 2$ is
proportional to the one of $\ell = 1$, that is, the one for the
first flux which, as notice above, multiplied by the mass $m$ is the
linear momentum density and then all are proportional to the linear
momentum relaxation time. On the other hand we do have that
%
\begin{equation}
\frac{\theta_{n_{\ell +1}}}{\theta_{n_{\ell}}} = \frac{\ell }{\ell
+1}\frac{5(1+x) + \ell -1}{5(1 + x) + \ell} \, ,
\end{equation}
for $\ell =1,2,3,\ldots$ and where $x=m/M$, then the ordered
sequence
%
\begin{equation}
\theta_{n_{1}} > \theta_{n_{2}} > \theta_{n_{3}} \ldots >
\theta_{n_{\ell}} > \theta_{n_{\ell +1}} > \ldots \, ,
\end{equation}
is verified, and it can be seen that $\theta_{n_{\ell}} \rightarrow
0$ as $\ell \rightarrow \infty$. Moreover, according to Eq. (23) it
follows that
%
\begin{equation}
\theta_{n_{\ell}} = \frac{5(1 + x)}{\ell \lbrack 5(1 + x) + \ell -
1]} \theta_{n_{1}} \, ,
\end{equation}

Comparing with the second flux ($\ell =2$) which is the one related
to the pressure tensor, it follows that for Brownian particles ($x
\gg 1$) $\theta_{n_{2}} \simeq \theta_{n_{1}}/2$ and for Lorentz
particles ($x\ll 1$) $\theta_{n_{2}} \simeq 5 \theta_{n_{1}}/12$. A
comparison with the third flux leads to the results that
$\theta_{n_{3}} \simeq \theta_{n_{1}}/3$ and $\theta_{n_{3}} \simeq
5 \theta_{n_{1}}/21$ for the Brownian and Lorentz particles
respectively. For any $\ell$ we do have approximately:

1) for the Brownian particle ($m/M \gg 1$)
%
\begin{equation}
\theta_{n_{\ell}} \simeq \frac{\theta_{n_{1}}}{\ell} \, \, ,
\end{equation}

2) for the Lorentz particle ($m/M\ll 1$)
%
\begin{equation}
\theta_{n_{\ell}} \simeq \frac{5}{(4 + \ell )\ell} \theta_{n_{1}} \,
\, ,
\end{equation}
or $\theta_{n_{\ell}} \simeq 5 \theta_{n_{1}}/\ell ^{2}$ for large
$\ell$.

According to Eq. (24) as the order of the flux largely increases its
Maxwell characteristic time approaches zero, and $\theta_{n_{\ell
+1}}/ \theta_{n_{\ell}}\simeq 1$, with both practically null. In
Fig. 1 it is displayed the ratio of some characteristic times
compared with the one of the linear momentum in terms of the ratio
$m/M$.

\begin{figure}[h]
\center
\includegraphics[width=10cm]{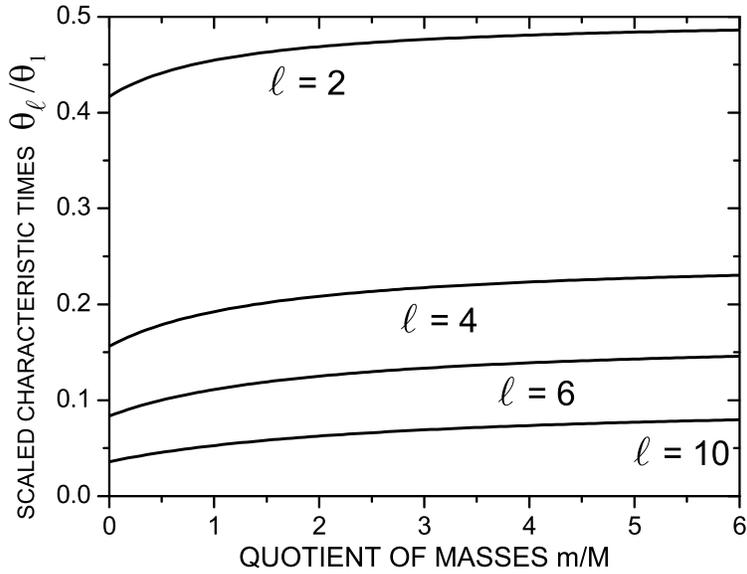}
\caption{The quotient between Maxwell characteristic times and the
one of the first flux as a function of $x=m/M$.}
\end{figure}

We emphasized that the hierarchy shown in Eq. (25) is evidencing the
important fact of the ever decreasing value of the Maxwell times
with the increasing order of the fluxes. This is quite relevant for
the question of how to introduce a contracted description (i.e.,
keeping a reduced number of fluxes) for characterizing the
particular type of hydrodynamic motion (diffusion, damped waves,
etc.) that the system displays, as shown later on.

\newpage
\section{Maxwell Times in Thermal Transport}

In this section we concentrate the attention on the evolution of the
thermal quantities
%
\begin{equation}
\{h(\mathbf{r},t); \, \mathbf{I}_{h}(\mathbf{r},t); \,
\{I_{h}^{[\ell ]}(\mathbf{r},t)\}\} \, ,
\end{equation}
with $\ell = 2, 3, \ldots$, that is, the density of energy and its
fluxes of all orders given in Eq. (11). Their evolution equations
are
%
\begin{equation}
\frac{\partial }{\partial t}I_{h}^{[\ell ]}(\mathbf{r},t) = \int
d^{3}p\frac{ p^{2}}{2m}\mathbf{u}^{[\ell
]}(\mathbf{p})\frac{\partial }{\partial t} f_{1}(
\mathbf{r},\mathbf{p};t) \, ,
\end{equation}
in which we need to introduce the kinetic equation for the single
particle $f_{1}(\mathbf{r},\mathbf{p};t)$ which is given in Ref.
[13]. Performing the lengthy calculations involved we finally arrive
at the general evolution equations ($\ell = 0,1,2, \ldots$)
%
\begin{equation*}
\frac{\partial }{\partial t}I_{h}^{[\ell]}(\mathbf{r},t) + \nabla
\cdot I_{h}^{[\ell +1]}(\mathbf{r},t) =
\end{equation*}
%
\begin{eqnarray}
&=& - \mathcal{F}(\mathbf{r},t)I_{n}^{[\ell +1]}(\mathbf{r},t) -
\sum \limits_{s=1}^{\ell}\sigma
(1,s)[\mathcal{F}(\mathbf{r},t)I_{h}^{[\ell -1]}(\mathbf{r},t)] +
\theta_{h_{\ell}}^{-1} I_{h}^{[\ell]}(\mathbf{r},t) +  \notag \\
&& a_{L0} \sum\limits_{s=1}^{\ell }\sigma (1,s)[\nabla I_{n}^{[\ell
-1]}(\mathbf{r},t)]+2(\ell +3)a_{L1}\nabla \cdot
I_{h}^{[\ell+1]}(\mathbf{r},t) +
\notag \\
&& S_{h}^{[\ell]}(\mathbf{r},t) \, ,
\end{eqnarray}
where
%
\begin{equation}
\theta_{h_{\ell}}^{-1} = (\ell + 2) |a_{\tau 0}| + [\ell^{2} +
5(\ell + 1)] |b_{\tau 1}| \, ,
\end{equation}
is the reciprocal of the Maxwell time associated to the $\ell$-th
order flux of energy. Moreover,
%
\begin{eqnarray}
S_{h}^{[\ell ]}(\mathbf{r},t) &=& (2\ell + 3) b_{\tau
0}\frac{m}{2}I_{n}^{[\ell
]}(\mathbf{r},t)+  \notag \\
&&b_{\tau 0}\{\hat{\sigma}_{\ell }[1^{[2]}I_{h}^{[\ell
-2]}(\mathbf{r},t)\}+b_{\tau 1}\frac{2}{m}\{\hat{\sigma}_{\ell
}[1^{[2]}I_{h2}^{[\ell -2]}(\mathbf{r},t)]\}+  \notag \\
&&a_{L0}m\nabla \cdot I_{n}^{[\ell +1]}(\mathbf{r},t) +  \notag \\
&&a_{L1}\frac{2}{m}\sum\limits_{s=1}^{\ell }\sigma (1,s)[\nabla
I_{h2}^{[\ell -1]}(\mathbf{r},t)]+  \notag \\
&&3(\ell +2)a_{\tau 1}\frac{2}{m}I_{h2}^{[\ell ]}(\mathbf{r},t) +  \notag \\
&&R_{h}^{[\ell ]}(\mathbf{r},t) \, ,
\end{eqnarray}
is a contribution which couples the thermal motion to the material
motion (thermal-striction effect), $R_{h}^{[\ell]}(\mathbf{r},t)$
contains contribution of all the other higher-order fluxes
($\geqslant \ell +2$). The several kinetic coefficients are
presented in Appendix B in Ref. [13].

Taking into account that
%
\begin{equation}
|b_{\tau 1}| = \frac{1}{5(1 + x)}|a_{\tau 0}| \, ,
\end{equation}
it follows that Maxwell times have the property that
%
\begin{equation}
\frac{\theta_{h_{\ell + 1}}}{\theta_{h_{\ell}}} = \frac{5(\ell +
2)(1 + x) + \ell^{2} + 5(\ell + 1)}{5(\ell +3)(1+x)+(\ell
+1)^{2}+5(\ell +2)} < 1 \, ,
\end{equation}
for $\ell = 0,1,2,3,\ldots$, and where $x=m/M$. The ordered sequence
%
\begin{equation}
\theta_{h_{0}} > \theta_{h_{1}} > \theta_{h_{2}} > \theta_{h_{3}} >
\ldots > \theta_{h_{\ell}} > \theta_{h_{\ell + 1}} > \ldots
\end{equation}
is verified in an analogous condition on the Maxwell times
associated to the particle motion. It may be noticed that for large
$\ell $, $\theta_{\ell}$ goes as $\ell^{2}$, i.e., going to zero as
$\ell \rightarrow \infty$.

For the Brownian particle ($x\gg 1$) the ratio in Eq. (35) tends
asymptotically to $(\ell + 2)/(\ell + 3)$ while for Lorentz
particles ($x\ll 1 $) the ratio tends to $[\ell^{2} + 5(2\ell +
3)]/[(\ell +1)^{2} + 5(2\ell + 5)]$]. Considering that the Maxwell
time of the first flux of mater, $\theta_{n_{1}}$ is $|a_{\tau
0}|^{-1}$ (cf. Eq. (22)) we can establish the relation between it
and the Maxwell time for the energy
%
\begin{equation}
\theta_{h_{0}} = \frac{1 + x}{2x + 3} \theta_{n_{1}} \, ,
\end{equation}
what tell us that $\theta_{h_{0}} < \theta_{n_{1}}$. For a Brownian
particle $\theta_{h_{0}} = \theta_{n_{1}}/2$ and for a Lorentz
particle $\theta_{h_{0}} = \theta_{n_{1}}/3$. Moreover, for general
$\ell$ we have that
%
\begin{equation}
\theta_{h_{\ell}} = \frac{5(1 + x)}{5(\ell + 2)(1 + x) + \ell^{2} +
5(\ell + 1)} \theta_{n_{1}} \, ,
\end{equation}

Therefore, recalling that the first flux of matter multiplied by the
particle mass is the linear momentum density, as a general rule we
can state that the energy relaxation time is smaller than momentum
relaxation time. As already notice at the end of the previous
Section, the hierarchy shown in Eq. (36) is quite relevant for the
choose of the contracted description to be used.

We proceed next to consider the case of a MHT at a quantum
mechanical level for the cases of phonons and electrons in
semiconductors.

\newpage
\section{Maxwell Times in MHT of Phonons in Semiconductors}

We consider a system of longitudinal acoustic \textsc{la} phonons in
a semiconductor in anharmonic interaction with the accompanying
transverse acoustic \textsc{ta} phonons. The sample is in contact
with a thermostat at temperature $T_{0}$. An external pumping source
drives the \textsc{la} phonon system out of equilibrium. The system
is characterized at the microscopic level by the Hamiltonian
%
\begin{equation}
\hat{H}=\hat{H}_{OS}+\hat{H}_{OB}+\hat{H}_{SB}+\hat{H}_{SP}\,,
\end{equation}
which consists of the Hamiltonian of the free \textsc{la} phonons
%
\begin{equation}
\hat{H}_{OS} = \sum \limits_{\mathbf{q}}\hbar \omega
_{\mathbf{q}}(a_{\mathbf{q}}^{\dag }a_{\mathbf{q}}+1/2) \, ,
\end{equation}
where $\omega_{\mathbf{q}}$ is the frequency dispersion relation and
the sum on $\mathbf{q}$ runs over the Brillouin zone, and the one of
the \textsc{ta} phonons, which we call the thermal bath in which the
\textsc{la} phonons are embedded, given by
%
\begin{equation}
\hat{H}_{0B} = \sum \limits_{\mathbf{q}} \hbar \Omega
_{\mathbf{q}}(b_{\mathbf{q}}^{\dag }b_{\mathbf{q}}+1/2) \, ,
\end{equation}
where $\Omega_{\mathbf{q}}$ is the frequency dispersion relation;
$a_{\mathbf{q}}^{\dag}$($a_{\mathbf{q}}$) and
$b_{\mathbf{q}}^{\dag}$($b_{\mathbf{q}}$) are the corresponding
creation (annihilation) operators in mode $\mathbf{q}$.

Moreover, the interaction of the \textsc{la} phonons with the
thermal bath is given by
%
\begin{equation}
\hat{H}_{SB} =
\sum\limits_{\mathbf{k},\mathbf{q}}M_{\mathbf{k},\mathbf{q}}a_{
\mathbf{q}}b_{\mathbf{k}+\mathbf{q}}^{\dag }b_{-\mathbf{k}}^{\dag
}+\mathrm{hc}\,,
\end{equation}
where we have retained the only process that contributes to the
kinetic equations (\textsc{la}$\leftrightarrows
$\textsc{ta}$+$\textsc{ta}) and we have neglected nonlinear
contributions; $M_{\mathbf{k},\mathbf{q}}$ accounts for the coupling
strengths. Finally $\hat{H}_{SP}$ is the interaction energy operator
for the phonons and an external pumping source, to be specified in
each case.

At the macroscopic level (the nonequilibrium thermodynamic state)
NESEF requires first to specify the basic variables that are to be
used to characterize the nonequilibrium ensemble [7-10]. A priori,
when the system is initially driven away from equilibrium, it is
necessary to include all the observables of the system what is
attained introducing many-particle dynamical operators [14,15]; in
the present case it suffices to take only the single-phonon
dynamical operators
$\hat{\nu}_{\mathbf{q},\mathbf{Q}}=a_{\mathbf{q}+\mathbf{
Q}/2}^{\dag }a_{\mathbf{q}-\mathbf{Q}/2}$ in the second-quantization
representation in reciprocal space. The two-phonon dynamical
operator and higher-order ones can be ignored because of
Bogoliubov's principle of correlation weakening [15]. Moreover, it
would be necessary to also include the amplitudes
$a_{\mathbf{q}}^{\dag}$ and $a_{\mathbf{q}}$ because their
eigenstates are the coherent states [16], and the phonon pairs [17],
both are disregarded because they are of no practical relevance for
the problem considered here. Another basic microdynamical variable
that needs be incorporated is the energy of the bath, and then the
basic set is composed of
%
\begin{equation}
\{\hat{\nu}_{\mathbf{q}}, \, \hat{\nu}_{\mathbf{q,Q}}, \,
\hat{H}_{OB}\}\,,
\end{equation}
with $\mathbf{Q}\neq 0$, where we have separated out the so-called
populations, $\hat{\nu}_{\mathbf{q}} = a_{\mathbf{q}}^{\dag
}a_{\mathbf{q}}$, from those with $\mathbf{Q}\neq 0$ which are
related to the change in space of the populations (they are also
called coherences [18]).

Therefore, for the present case the nonequilibrium statistical
operator is
%
\begin{equation}
\varrho_{\epsilon}(t) = \exp \{-\hat{S}(t,0) + \int
\limits_{-\infty}^{t} dt^{\prime} e^{\epsilon (t^{\prime }-t)}
\frac{d}{dt^{\prime}} \hat{S}(t^{\prime},t^{\prime}-t)\} \, ,
\end{equation}
where
%
\begin{eqnarray}
\hat{S}(t,0) &=& - \ln \bar{\varrho}(t,0)=  \notag \\
&=& \phi (t) + \sum \limits_{\mathbf{q}}F_{\mathbf{q}}(t)a_{q}^{\dag
}a_{q} + \sum \limits_{\mathbf{qQ}\neq 0} F_{\mathbf{qQ}}(t)
a_{q+Q/2}^{\dag} a_{q-Q/2} + \beta _{0}\hat{H}_{0B} \, ,
\end{eqnarray}
%
\begin{equation}
\hat{S}(t^{\prime},t^{\prime }-t) = \exp \left\{ -\frac{1}{i\hbar
}(t^{\prime }-t) \hat{H} \right\} \hat{S}(t^{\prime},0) \exp \left\{
\frac{1}{i\hbar} (t^{\prime}-t) \hat{H}\right\} \, ,
\end{equation}
$\hat{H}_{0B}$ is given in Eq. (41), $\phi (t) = \ln \bar{Z}(t)$
plays the role of a logarithm of a nonequilibrium partition function
$\bar{Z}(t)$, and $\beta_{0} = 1/(k_{B}T_{0})$.

The average values of the microdynamical variables in set (43) over
the nonequilibrium ensemble provide the variables that characterized
the nonequilibrium macroscopic state of the system, which we
indicate by
%
\begin{equation}
\{\nu _{\mathbf{q}}(t), \, \nu_{\mathbf{q}\mathbf{Q}}(t), \, E_{B}\}
\, ,
\end{equation}
and the nonequilibrium thermodynamic variables conjugated to them
are
%
\begin{equation}
\{F_{\mathbf{q}}(t), \, F_{\mathbf{q}\mathbf{Q}}(t), \, \beta _{0}\}
\, .
\end{equation}

Going over to direct space (anti-transforming Fourier in variable
$\mathbf{Q}$) we obtain the space and crystalline momentum-dependent
distribution function $\nu_{\mathbf{q}}(\mathbf{r},t)$, in terms of
which is built the phonon Higher-Order Generalized
Hydro-Thermodynamics which consists of two families of
hydrodynamical variables, namely, the one associated to the
quasi-particles (the phonons) motion (we call it the $n$-family)
%
\begin{equation}
\{n(\mathbf{r},t), \, \mathbf{I}_{n}(\mathbf{r},t), \,
\{I_{n}^{[\ell]}(\mathbf{r},t)\}\} \, ,
\end{equation}
where $\ell =2,3,\ldots $, and $n(\mathbf{r},t)$ stands for the
number of phonons at time $t$ in position $\mathbf{r}$, namely
%
\begin{equation}
n(\mathbf{r},t) = \sum \limits_{\mathbf{q}}\nu
_{\mathbf{q}}(\mathbf{r},t) \, ,
\end{equation}
the first flux of this quantity
%
\begin{equation}
\mathbf{I}_{n}(\mathbf{r},t) = \sum \limits_{\mathbf{q}} \nabla
_{\mathbf{q}} \omega _{\mathbf{q}} \, \nu_{\mathbf{q}}(\mathbf{r},t)
\, ,
\end{equation}
and the higher-order fluxes
%
\begin{equation}
I_{n}^{[\ell ]}(\mathbf{r},t) = \sum\limits_{\mathbf{q}}u^{[\ell
]}(\mathbf{q}) \, \nu_{\mathbf{q}}(\mathbf{r},t) \, ,
\end{equation}
with $\ell = 2,3,\ldots $, defining the $\ell$-order flux ($\ell
$-rank tensor) where
%
\begin{equation}
u^{[\ell ]}(\mathbf{q}) = [\nabla _{\mathbf{q}}\omega
_{\mathbf{q}}:\ldots \ell - times \ldots :\nabla _{\mathbf{q}}\omega
_{\mathbf{q}}]\,,
\end{equation}
is a $\ell$-rank tensor consisting of the tensorial inner product of
$\ell $-times the group velocity of the $\mathbf{q}$-mode phonon,
$\nabla_{\mathbf{q}}\omega_{\mathbf{q}}$.

On the other hand we do have the family associated to the energy
motion (heat transport, and we call it the $h$-family)
%
\begin{equation}
\{h(\mathbf{r},t), \, \mathbf{I}_{h}(\mathbf{r},t), \,
\{I_{h}^{[\ell ]}(\mathbf{r},t)\}\} \, ,
\end{equation}
where
%
\begin{equation}
h(\mathbf{r},t)=\sum\limits_{\mathbf{q}}\hbar \omega
_{\mathbf{q}}\nu _{\mathbf{q}}(\mathbf{r},t)\,,
\end{equation}
%
\begin{equation}
\mathbf{I}_{h}(\mathbf{r},t) = \sum \limits_{\mathbf{q}} \hbar
\omega_{\mathbf{q }} \nabla _{\mathbf{q}} \omega_{\mathbf{q}} \,
\nu_{\mathbf{q}}(\mathbf{r},t) \, ,
\end{equation}
%
\begin{equation}
I_{h}^{[\ell ]}(\mathbf{r},t) = \sum \limits_{\mathbf{q}}\hbar
\omega_{\mathbf{q}}u^{[\ell ]}(\mathbf{q}) \,
\nu_{\mathbf{q}}(\mathbf{r},t) \, ,
\end{equation}
which are, respectively, the energy density, its first (vectorial)
flux, and the higher-order ($\ell = 2,3,\ldots$) tensorial fluxes at
time $t$ in position $\mathbf{r}$.

Consequently, the hydrodynamic equations of motion (evolution
equations for the quantities above) are
%
\begin{equation}
\frac{\partial }{\partial t}I_{p}^{[\ell ]}(\mathbf{r},t) = \sum
\limits_{ \mathbf{q}}K_{p}^{[\ell]}(\mathbf{q}) \frac{\partial
}{\partial t} \nu_{\mathbf{q}}(\mathbf{r},t) \, ,
\end{equation}
where, for $p \equiv n$, $K_{n}^{[\ell]}(\mathbf{q}) = u^{[\ell
]}(\mathbf{q})$ and for $p \equiv h$, $K_{h}^{[\ell ]}(\mathbf{q}) =
\hbar \omega_{\mathbf{q}}u^{[\ell]}(\mathbf{q})$, recalling that
$\ell = 0$ corresponds to the densities, $\ell = 1$ to the vectorial
fluxes, and $\ell \geqslant 2$ for the higher-order fluxes.

Evidently, all the evolution equations in Eq. (58) are determined by
the unique equation of motion for the single-phonon distribution
function $\nu _{\mathbf{q}}(\mathbf{r},t)$. It follows from the
evolution equation which results in taking the average over the
nonequilibrium ensemble of the quantum mechanical Heisenberg
equation of motion for the microdynamical variable
$\hat{\nu}_{\mathbf{q}\mathbf{Q}} =
a_{\mathbf{q}+\mathbf{Q}/2}^{\dag} a_{\mathbf{q}-\mathbf{Q}/2}$,
(for practical convenience is calculated in reciprocal space), that
is
%
\begin{equation}
\frac{\partial }{\partial t} \nu _{\mathbf{q,Q}}(t) = \mathrm{Tr}
\left\{ \frac{1}{i\hbar }[\hat{\nu}_{\mathbf{q}\mathbf{Q}},\hat{H}]
\varrho_{\epsilon}(t) \times \varrho_{B} \right\} \, .
\end{equation}
where $\varrho_{\epsilon}(t)$ is the nonequilibrium statistical
operator of Eq. (44), and $\varrho_{B}$ is the canonical statistical
distribution of the thermal bath (the \textsc{ta} phonons) at the
temperature $T_{0}$. Performing such average is extremely difficult
and then it is necessary to resort to the introduction of a
practical nonlinear quantum kinetic theory [7-10,19,20], which, for
the present case, can be used in the approximation consisting in
retaining in the collision integral contributions up to second order
in the interaction strength (Markovian approximation [9,19,21]). In
that approximation it follows, after going over to direct space, a
generalization of the so-called Peierls-Boltzmann equation given by
[22]
%
\begin{eqnarray}
\frac{\partial }{\partial t} \nu _{\mathbf{q}}(\mathbf{r},t) &=& -
\nabla \cdot \nabla_{\mathbf{q}} \tilde{\omega}_{\mathbf{q}}
\nu_{\mathbf{q}}(\mathbf{r},t) - \Gamma_{\mathbf{q}} \,
[\nu_{\mathbf{q}}(\mathbf{r},t) - \nu_{\mathbf{q}}^{0}] \nonumber \\
&& + J_{\mathbf{q}S}(\mathbf{r},t) \, ,
\end{eqnarray}
where $J_{\mathbf{q}S}$ contains the effect of the presence of
external sources, to be specified in each particular case, and
\begin{equation*}
\nu_{\mathbf{q}}^{0} = \frac{1}{ e^{ \beta_{0} \hbar
\omega_{\mathbf{q}} } -1} \, ,
\end{equation*}
is the distribution in equilibrium,
%
\begin{equation}
\tilde{\omega}_{\mathbf{q}} = \omega_{\mathbf{q}} + P_{\mathbf{q}}
\, ,
\end{equation}
with
%
\begin{equation}
P_{\mathbf{q}} = \frac{\pi}{\hbar ^{2}} \sum
\limits_{\mathbf{k}}|M_{\mathbf{kq} }|^{2} \frac{1 + \nu
_{\mathbf{k}}^{TA} + \nu_{\mathbf{k}+\mathbf{q}}^{TA}}{\Omega
_{\mathbf{k} + \mathbf{q}} + \Omega_{\mathbf{k}} - \omega
_{\mathbf{q}}} \, ,
\end{equation}
which is the so-called self-energy correction, and
%
\begin{equation}
\Gamma _{\mathbf{q}} = \frac{\pi }{\hbar^{2}} \sum
\limits_{\mathbf{k}} |M_{ \mathbf{kq}}|^{2} (1 + \nu
_{\mathbf{k}}^{TA} + \nu_{\mathbf{k}+\mathbf{q}}^{TA}) \delta
(\Omega_{\mathbf{k}+\mathbf{q}} + \Omega_{\mathbf{k}} -
\omega_{\mathbf{q}}) + \overline{\Gamma}_{\mathbf{q}} \, ,
\end{equation}
plays the role of the reciprocal of a relaxation time per mode, say
$\tau_{\mathbf{q}}^{-1}$, (towards the equilibrium distribution),
where
%
\begin{equation}
\nu_{\mathbf{k}}^{TA} = \frac{1}{e^{\beta_{0} \hbar
\Omega_{\mathbf{k} }} - 1} \, ,
\end{equation}
is the distribution in equilibrium at temperature $T_{0}$ of the
\textsc{ta} phonons. In this Eq. (63) we have written explicitly the
known contribution due to the anharmonic interaction, and in a
Mathiessen rule [23] are also indicated in $\overline{\Gamma}
_{\mathbf{q}}$ the sum of the inverse of the relaxation times due to
the presence of impurities, imperfections, boundary conditions, etc.

Using Eq. (60) in Eq. (58), we obtain for the set of evolution
equations that
%
\begin{eqnarray}
\frac{\partial }{\partial t}I_{p}^{[\ell ]}(\mathbf{r},t) & = & \sum
\limits_{\mathbf{q}} K_{p}^{[\ell]}(\mathbf{q}) [- \nabla \cdot
\nabla_{\mathbf{q}} \tilde{\omega}_{\mathbf{q}}] \nu_{\mathbf{q}}(\mathbf{r},t) \nonumber \\
&& - \Gamma_{\mathbf{q}} \, [\nu_{\mathbf{q}}(\mathbf{r},t) -
\nu_{\mathbf{q}}^{0}] + J_{pS}^{[\ell]}(\mathbf{r},t) \, ,
\end{eqnarray}
for both families of sets (49) and (54), $p \equiv n$ and $p \equiv
h$ respectively. A closure for the set of these equations must be
introduced, that is, to express $\nu_{\mathbf{q}}(\mathbf{r},t)$
that appears on the right hand side in terms of the hydrodynamic
variables in the sets of Eqs. (49) and (54). First it can be noticed
that Eq. (65) involves an enormous set of coupled nonlinear
integro-differential equations, \emph{i.e.}, both densities together
with their fluxes of all orders are coupled through all these
equations. To proceed further it is necessary to introduce a
\emph{contraction of description}, that is to say, to reduce the
number of variables to be used, as described elsewhere [9,10,24].
Moreover, the two families in the sets of Eqs. (49) and (54) are
coupled by cross-terms that account for thermo-striction effects. In
cases where these effects are not particularly relevant they can be
disregarded and we obtain two independent sets of evolution
equations, one for the $n$-family and the other for the $h$-family.

Admitting that the just mentioned conditions are well satisfied, we
introduce a contracted description of first order, and proceed to
analyze the evolution equations for the energy density and its first
flux. The resulting equations are [22]
%
\begin{eqnarray}
\frac{\partial }{\partial t}h(\mathbf{r},t) + \nabla \cdot
\mathbf{I}_{h}(\mathbf{r},t) = a_{L}(t)\nabla \cdot
\mathbf{I}_{h}(\mathbf{r},t) \nonumber \\
 - \theta _{h}^{-1}(t)h(\mathbf{r},t) + J_{hS}(\mathbf{r},t)\,,
\end{eqnarray}
%
\begin{eqnarray}
\frac{\partial }{\partial t}\mathbf{I}_{h}(\mathbf{r},t) + \nabla
\cdot I_{h}^{[2]}(\mathbf{r},t) = b_{L}^{[2]}(t) \cdot \nabla
h(\mathbf{r},t) \nonumber \\
 - \theta_{I}^{-1}(t) \mathbf{I}_{h}(\mathbf{r},t) +
\mathbf{J}_{IS}(\mathbf{r},t) \, ,
\end{eqnarray}
where
%
\begin{equation}
a_{L}(t) = \sum \limits_{\mathbf{q}} \hbar \omega_{\mathbf{q}}
\mathbf{a}_{I}(\mathbf{q},t) \cdot \nabla_{\mathbf{q}}
P_{\mathbf{q}} \, ,
\end{equation}
%
\begin{equation}
b_{L}^{[2]}(t) = \sum \limits_{\mathbf{q}} \hbar \omega_{\mathbf{q}}
a_{h}(\mathbf{q},t) [\nabla_{\mathbf{q}} P_{\mathbf{q}} \colon
\nabla_{\mathbf{q}} \omega_{\mathbf{q}}] \, ,
\end{equation}
%
%
\begin{equation}
a_{h}(\mathbf{q},t) = \hbar \omega_{\mathbf{q}} \nu_{\mathbf{q}}(t)
[1 + \nu_{ \mathbf{q}}(t)] \Big{[} \sum \limits_{\mathbf{q}} (\hbar
\omega_{\mathbf{q}})^{2} \nu_{\mathbf{q}}(t)[1 +
\nu_{\mathbf{q}}(t)] \Big{]}^{-1} ,
\end{equation}
%
\begin{eqnarray}
\mathbf{a}_{I}(\mathbf{q},t) &=& \hbar \omega_{\mathbf{q}}
\nu_{\mathbf{q}}(t) [1 + \nu_{\mathbf{q}}(t)] \Big{[} \sum
\limits_{\mathbf{q}} (\hbar
\omega_{\mathbf{q}})^{2} \nu_{\mathbf{q}}(t) [1 + \nu_{\mathbf{q}}(t)]  \times  \notag \\
&& \lbrack \nabla_{\mathbf{q}} \omega_{\mathbf{q}} \cdot
\nabla_{\mathbf{q}} \omega_{\mathbf{q}}] \Big{]}^{-1} \nabla
_{\mathbf{q}} \omega_{\mathbf{q}} \, ,
\end{eqnarray}
$P_{\mathbf{q}}$ and $\Gamma_{\mathbf{q}}$ are given in Eqs. (62)
and (63), and
%
\begin{equation}
\nu _{\mathbf{q}}(t) = \frac{1}{e^{F_{\mathbf{q}}(t)} - 1}  \, ,
\end{equation}
with $F_{\mathbf{q}}$ of Eq. (48), we recall that
%
\begin{equation}
I_{h}^{[2]}(\mathbf{Q},t) = \sum \limits_{\mathbf{q}} \hbar
\omega_{\mathbf{q}} [\nabla_{\mathbf{q}} \omega_{\mathbf{q}} \colon
\nabla_{\mathbf{q}} \omega_{\mathbf{q}}] \nu_{\mathbf{qQ}}(t) \, ,
\end{equation}
and $J_{hS}$ and $\mathbf{J}_{IS}$ are the contributions due to the
coupling with the external sources that are present which are to be
specified in each case being considered. We recall that $[\cdots \,
\colon \, \cdots]$ stands for inner tensorial product.

Finally,
%
\begin{equation}
\theta _{h}^{-1}(t) = \sum \limits_{\mathbf{q}} w_{h}(\mathbf{q},t)
\Gamma_{\mathbf{q}} \, ,
\end{equation}
and
%
\begin{equation}
\theta_{I}^{-1}(t) = \sum \limits_{\mathbf{q}} w_{I}(\mathbf{q},t)
\Gamma_{\mathbf{q}} \, ,
\end{equation}
with
%
\begin{equation}
w_{h}(\mathbf{q},t) = \frac{(\hbar \omega_{\mathbf{q}})^{2}\nu
_{\mathbf{q}} (1 +
\nu_{\mathbf{q}})}{\sum\limits_{\mathbf{q}^{\prime }}(\hbar \omega
_{\mathbf{q}^{\prime }})^{2} \nu_{\mathbf{q}^{\prime}} (1 +
\nu_{\mathbf{q}^{\prime}})} \, ,
\end{equation}
%
\begin{equation}
w_{I}(\mathbf{q},t) = \frac{(\hbar \omega_{\mathbf{q}})^{2}
\nu_{\mathbf{q}}(1 + \nu_{\mathbf{q}})}{\sum
\limits_{\mathbf{q}^{\prime}}(\hbar \omega
_{\mathbf{q}^{\prime}})^{2} \nu_{\mathbf{q}^{\prime}}(1 + \nu
_{\mathbf{q}^{\prime }})} \frac{|\nabla_{\mathbf{q}} \omega
_{\mathbf{q}}|^{2}}{|\nabla_{\mathbf{q}^{\prime}}
\omega_{\mathbf{q}^{\prime}}|^{2}} \, ,
\end{equation}
which have the property of normalization, that is, it is verified
that
%
\begin{equation}
\sum \limits_{\mathbf{q}} w_{h}(\mathbf{q},t) = 1 \, ,
\end{equation}
and
%
\begin{equation}
\sum \limits_{\mathbf{q}} w_{I}(\mathbf{q},t) = 1 \, ,
\end{equation}

Equations (74) and (75) define the reciprocal of Maxwell times
associated to energy and its first flux, respectively, in this MHT
of order 1. We can see that each Maxwell time follows a kind of
Mathiessen rule: its inverse is a superposition of the reciprocal of
the relaxation times of each mode $\tau_{\mathbf{q}}^{-1} =
\Gamma_{\mathbf{q}}$, multiplied by a weighting function
$w(\mathbf{q},t)$. Because of the normalization condition (Eqs. (78)
and (79)) it can be considered each $w(\mathbf{q},t)$ as the
probability at each time $t$ of the contribution of the
mode-relaxation time $\tau _{\mathbf{q}}$.

In a Debye model both Maxwell times are equal, \emph{i.e.},
%
\begin{equation}
\theta_{h}^{-1}(t) = \theta _{I}^{-1}(t) = \frac{\sum
\limits_{\mathbf{q}}(sq)^{2} \nu_{\mathbf{q}}(1 + \nu_{\mathbf{q}})
\Gamma_{\mathbf{q}}}{\sum \limits_{\mathbf{q}}(sq)^{2}
\nu_{\mathbf{q}}(1 + \nu_{\mathbf{q}})}
\end{equation}
and we recall that $q \leq q_{Debye}$.

Taking into account that $|\nabla_{\mathbf{q}} \omega_{\mathbf{q}}|
\simeq s$ for small values of $q$ and that $|\nabla_{\mathbf{q}}
\omega_{\mathbf{q}}| < s$ for intermediate to large values of $q$,
we can estimate that in general $\theta_{h} > \theta_{I}$. It is
conjectured that we would do have a hierarchy
\begin{equation*}
\theta_{h} > \theta_{I} > \theta_{I_{2}} > \ldots \, ,
\end{equation*}
quite similar to the case of the classical fluid of previous
sections.

\newpage
\section{Maxwell Times in MHT of Electrons in Semiconductors}

We consider the case of a n-doped polar semiconductor, e.g., n-GaAs
or n-GaN, whose Hamiltonian quantum mechanical operator is
%
\begin{equation}
\hat{H} = \hat{H}_{0e} + \hat{H}_{eph} + \hat{H}_{0ph} +
\hat{H}_{e\varphi} \, ,
\end{equation}
where $\hat{H}_{0e}$ is the Hamiltonian of the carriers (a
concentration $n$ of electrons in conduction band) taken in the
effective mass approximation, with energy dispersion relation
$\epsilon _{\mathbf{k}}=\hbar ^{2}k^{2}/2m^{\ast}$. $\hat{H}_{eph}$
is the Hamiltonian of the interactions of carriers and optical and
acoustical phonons: we consider, in this case of polar
semiconductors only the polar (Fr\"{o}hlich) interaction with
\textsc{lo} phonons, which is by large the predominant process of
relaxation of the carriers to the lattice. $\hat{H}_{0ph}$ is the
Hamiltonian of the free \textsc{lo} phonons. $\hat{H}_{e\varphi}$
accounts for the interaction of the carriers with external pumping
sources. We assume that the \textsc{lo} phonons remain in
equilibrium with an external reservoir at temperature $T_{0}$. The
polarization effects (due the Coulomb interaction) modifying the
mean field potential of the band states is not included, that is, we
disregard the contribution from the plasmon states.

Indicating by $c_{\mathbf{k}}(c_{\mathbf{k}}^{\dag})$ the creation
and annihilation operators in band states $\mathbf{k}$ (we omit the
spin index), as in the case of phonons of the previous Section we
introduce as basic microdynamical variables for the corresponding
NESEF the quantities
%
\begin{equation}
\{\hat{n}_{\mathbf{kQ}} = c_{\mathbf{k}+\mathbf{Q}/2}^{\dag}
c_{\mathbf{k}-\mathbf{Q}/2}; \, \hat{H}_{0ph}\} \, .
\end{equation}

For $Q=0$, $\hat{n}_{\mathbf{k}} = c_{\mathbf{k}}^{\dag}
c_{\mathbf{k}}$ is the operator occupation number in state
$\mathbf{k}$, and the others with $\mathbf{Q} \neq 0$ accounts for
variations in space. For the \textsc{lo} phonons we have taken
$\hat{H}_{0ph}$ once, as said, we consider them in constant
equilibrium at temperature $T_{0}$ and therefore described by a
canonical distribution. Hence, the statistical operator in NESEF of
the electrons is
%
\begin{equation}
\varrho _{\epsilon }(t) = \exp \Big\{ - \hat{S}(t,0) + \int
\limits_{- \infty}^{t} dt^{\prime} e^{\epsilon (t^{\prime }-t)}
\hat{S}(t^{\prime },t^{\prime }-t) \Big\} \, ,
\end{equation}
where
%
\begin{eqnarray}
\hat{S}(t_{1},t_{2}) &=& \ln \bar{\varrho}(t_{1},t_{2}) =  \notag \\
&=& \phi (t_{1}) + \sum \limits_{\mathbf{k}} [F_{\mathbf{k}}(t_{1})
\hat{n}_{\mathbf{k}}(t_{2}) + \sum \limits_{\mathbf{Q} \neq 0}
F_{\mathbf{kQ}}(t_{1}) \hat{n}_{\mathbf{kQ}}(t_{2})] \, ,
\end{eqnarray}
where we recall that $\phi$ ensures the normalization of both
$\bar{\varrho}$ and $\varrho_{\epsilon}$, and $F_{\mathbf{k}}$ and
$F_{\mathbf{kQ}}$ are the nonequilibrium thermodynamic variables
conjugated to the basic ones $\hat{n}_{\mathbf{k}}$ and
$\hat{n}_{\mathbf{kQ}}$ ($\mathbf{Q}\neq 0$). In Eq. (84) $t_{1}$
refers to the evolution in time of the nonequilibrium thermodynamic
variables and $t_{2}$ to the evolution in time of the dynamical
variables (Heisenberg representation).

Following the formalism we proceed to the derivation of the kinetic
equation for the variable
%
\begin{equation}
n_{\mathbf{kQ}}(t) = \mathrm{Tr} \{\hat{n}_{\mathbf{kQ}}
\varrho_{\epsilon }(t) \, .
\end{equation}

In the Markovian approximation after a calculation in reciprocal
space and next going over to direct space it follows for
%
\begin{equation}
f_{\mathbf{k}}(\mathbf{r},t) = \sum
\limits_{\mathbf{Q}}n_{\mathbf{kQ}}(t) e^{i \mathbf{Q} \cdot
\mathbf{r}} \, ,
\end{equation}
that
%
\begin{equation*}
\frac{\partial}{\partial t} f_{\mathbf{k}}(\mathbf{r},t) +
\frac{i}{\hbar} \nabla_{\mathbf{k}} \epsilon_{\mathbf{k}} \cdot
\nabla_{\mathbf{r}} f_{\mathbf{k}}(\mathbf{r},t) =
\end{equation*}
\begin{eqnarray}
&& \frac{2\pi}{\hbar} \sum \limits_{\mathbf{q}}
|\mathcal{C}_{\mathbf{q}}|^{2} \{[ \nu_{\mathbf{q}}(t) + 1]
f_{\mathbf{k+q}}(\mathbf{r},t) [1 -
f_{\mathbf{k}}(\mathbf{r},t)] -  \notag \\
&& \nu_{\mathbf{q}}(t) f_{\mathbf{k}}(\mathbf{r},t) [1 -
f_{\mathbf{k+q}}(\mathbf{r},t)]\} \delta (\epsilon_{\mathbf{k+q}} -
\epsilon_{\mathbf{k}} - \hbar \omega_{\mathbf{q}}) -  \notag \\
&& \frac{2\pi}{\hbar} \sum \limits_{\mathbf{q}}
|\mathcal{C}_{\mathbf{q}}|^{2} \{ [\nu_{\mathbf{q}}(t)
+ 1] f_{\mathbf{k}}(\mathbf{r},t) [1 - f_{\mathbf{k-q}}(\mathbf{r},t)] -  \notag \\
&& \nu_{\mathbf{q}}(t) f_{\mathbf{k-q}}(\mathbf{r},t) [1 -
f_{\mathbf{k}}(\mathbf{r},t)]\} \delta (\epsilon_{\mathbf{k-q}} -
\epsilon_{\mathbf{k}} + \hbar \omega_{\mathbf{q}})\} +  \notag \\
&& J_{S\mathbf{k}}(\mathbf{r},t) \, ,
\end{eqnarray}
where $\mathcal{C}_{\mathbf{q}}$ is the matrix element of the
Fr\"{o}hlich electron-\textsc{lo} phonon interaction, $\omega
_{\mathbf{q}}$ the frequency dispersion relation of the \textsc{lo}
phonons, we recall that $\epsilon_{\mathbf{k}} = \hbar
^{2}k^{2}/2m^{\ast}$, $J_{S\mathbf{k}}$ the interaction with
external sources to be specified in each case, and
$\nu_{\mathbf{q}}$ is the population of the \textsc{lo} phonons in
equilibrium at temperature $T_{0}$.

As in previous Sections we introduce the hydro-thermodynamic
variables
%
\begin{equation}
I_{p}^{[\ell]}(\mathbf{r},t) = \sum \limits_{\mathbf{k}}K_{p}^{[\ell
]}(\mathbf{k}) f_{\mathbf{k}}(\mathbf{r},t) \, ,
\end{equation}
%
\begin{equation}
K_{n}^{[\ell ]}(\mathbf{k}) = \left[ \frac{\hbar
\mathbf{k}}{m^{\ast}} \ldots \ell - times \ldots \frac{\hbar
\mathbf{k}}{m^{\ast}} \right] \, ,
\end{equation}
%
\begin{equation}
K_{h}^{[\ell ]}(\mathbf{k}) = \frac{\hbar^{2}k^{2}}{2m^{\ast}}
K_{n}^{[\ell ]}(\mathbf{k}) \, ,
\end{equation}
corresponding to the families of particle motion ($p \equiv n$) and
energy motion ($p \equiv h$). The evolution equations are evidently
given by
%
\begin{equation}
\frac{\partial }{\partial t}I_{p}^{[\ell ]}(\mathbf{r},t) = \sum
\limits_{\mathbf{k}}K_{p}^{[\ell ]}(\mathbf{k}) \frac{\partial
}{\partial t} f_{\mathbf{k}}(\mathbf{r},t) \, ,
\end{equation}

In what follows, for better visualization, we restrict the analysis
to a description of order 2 of the heat transport introducing the
reduced set of basic variables consisting of
%
\begin{equation}
\{ h(\mathbf{r},t); \, I_{h}(\mathbf{r},t); \, I_{h2}(\mathbf{r},t)
\} \, ,
\end{equation}
where
%
\begin{equation}
h(\mathbf{r},t) = \sum \limits_{\mathbf{k}}
\frac{\hbar^{2}k^{2}}{2m^{\ast}} f_{\mathbf{k}}(\mathbf{r},t) \, ,
\end{equation}
%
\begin{equation}
\mathbf{I}_{h}(\mathbf{r},t) = \sum \limits_{\mathbf{k}} \frac{\hbar
\mathbf{k}}{m^{\ast}} f_{\mathbf{k}}(\mathbf{r},t) \, ,
\end{equation}
and we separate the second order flux $I_{h}^{[2]}(\mathbf{r},t)$ in
the trace part and the remaining traceless part which is neglected.
The trace of $I^{[2]}$ is
%
\begin{equation}
I_{h2}(\mathbf{r},t) = \sum \limits_{\mathbf{k}} \left( \frac{\hbar
k}{m^{\ast}} \right)^{2} f_{\mathbf{k}}(\mathbf{r},t) \, .
\end{equation}

The auxiliary statistical operator is then given by
%
\begin{eqnarray}
\bar{\varrho}(t,0) &=& \exp \Big\{ - \phi (t,0)  \notag \\
&& - \sum \limits_{\mathbf{Q}} [ F_{h}(\mathbf{Q},t)
\hat{h}(\mathbf{Q}) + \boldsymbol{\varphi}_{h}(\mathbf{Q},t) \cdot \mathbf{I}_{h}(\mathbf{Q},t)   \notag \\
&& + F_{h2}(\mathbf{Q},t) I_{h2}(\mathbf{Q})] \Big\} \, ,
\end{eqnarray}
which follows from Eq. (84) after taking, to be consistent with the
contraction introduced, that
%
\begin{eqnarray}
F_{\mathbf{kQ}}(t) &=& \frac{\hbar^{2} k^{2}}{2m^{\ast}}
F_{h}(\mathbf{Q},t) + \frac{\hbar^{2} k^{2}}{2m^{\ast}} \hbar
\mathbf{k} \cdot \boldsymbol{\varphi}_{h}(\mathbf{Q},t)  +  \notag \\
&& \left( \frac{\hbar k}{m^{\ast}} \right)^{2} F_{h2}(\mathbf{Q},t)
\, .
\end{eqnarray}

The equation of evolution for these hydro-thermodynamic variables in
the absence of external sources are
%
\begin{equation}
\frac{\partial}{\partial t} h(\mathbf{r},t) = - \nabla \cdot
\mathbf{I}_{h}(\mathbf{r},t) - \theta_{h}^{-1} h(\mathbf{r},t) -
a_{0}I_{h2}(\mathbf{r},t) \, ,
\end{equation}
%
\begin{equation}
\frac{\partial }{\partial t}\mathbf{I}_{h}(\mathbf{r},t) = -\nabla
I_{h2}(\mathbf{r},t) - \theta_{I_{h}}^{-1}
\mathbf{I}_{h}(\mathbf{r},t) \, ,
\end{equation}
%
\begin{equation}
\frac{\partial}{\partial t} I_{h2}(\mathbf{r},t) = -
\frac{q}{m^{\ast}\beta} \nabla \cdot \mathbf{I}_{h}(\mathbf{r},t) -
a_{2} h(\mathbf{r},t) - \theta_{I_{h2}}^{-1} I_{h2}(\mathbf{r},t) \,
.
\end{equation}

The expression for the trace of the divergence of the third order
flux in terms of the basic macrovariables, namely
%
\begin{equation}
\mathrm{Tr} \{\nabla \cdot I_{h}^{[3]}(\mathbf{r},t)\} =
\frac{q}{m^{\ast} \beta} \nabla \cdot \mathbf{I}_{h}(\mathbf{r},t) -
a_{2}h(\mathbf{r},t) \, ,
\end{equation}
was obtained resorting to Heims-Jaynes perturbation expansion for
averages [25] around the homogeneous state where
%
\begin{equation}
a_{2} = \frac{1}{m^{\ast} \beta \theta_{I_{2}}} \, ,
\end{equation}
after using that
%
\begin{equation}
f_{\mathbf{k}} = \frac{1}{\exp \{\beta \hbar^{2} k^{2}/2m^{\ast} \}
+ 1} \, ,
\end{equation}
%
\begin{equation}
\beta = \frac{1}{k_{B} T^{\ast}} \, ,
\end{equation}
with $T^{\ast}$ being the nonequilibrium temperature
(quasitemperature) of the electrons, and in Eq. (84) making the
identification
%
\begin{equation}
F_{\mathbf{k}} \equiv \frac{\hbar^{2}k^{2}}{2m^{\ast}} \beta (t) \,
,
\end{equation}

In Eqs. (98) to (100) are present the Maxwell times $\theta_{h}$,
$\theta_{I_{h}}$ and $\theta_{I_{h2}}$, associated to the energy,
its vectorial (first) flux (current of heat) and the second flux
respectively. They are given by
%
\begin{equation}
\tau _{h}^{-1} = \frac{14}{5} \xi \tau_{0}^{-1} \, ,
\end{equation}
%
\begin{equation}
\tau _{I_{h}}^{-1} = \frac{2}{25} \xi \tau _{h}^{-1} =
\frac{2}{25}\frac{14}{5} \xi^{2} \tau_{0}^{-1} \, ,
\end{equation}
%
\begin{equation}
\tau_{I_{h2}}^{-1} = \frac{1}{28} \xi^{2} \tau_{h}^{-1} =
\frac{1}{25}\frac{14}{5} \xi ^{3} \tau_{0}^{-1} \, ,
\end{equation}
where
%
\begin{equation}
\xi = \beta \hbar \omega_{0} = \frac{\hbar
\omega_{0}}{k_{B}T^{\ast}} \, ,
\end{equation}
and
%
\begin{equation}
\tau_{0}^{-1} = \, ,
\end{equation}
in the calculation being used the parameters corresponding to GaAs
given in Table I.

%
\begin{table}[htbp]
\caption{Parameters characteristic of GaAs}
\label{tab1}
\begin{tabular}{cc}
\hline\hline
Parameter                                                & Value                \\
\tableline electron effective mass, $m^{\ast}$           & 0.067$m_{0}$         \\
static dielectric constant, $\varepsilon_{0}$            & 12.91                \\
optical dielectric constant, $\varepsilon_{\infty}$      & 10.91                \\
\textsc{lo} phonon energy, $\hbar \omega_{LO}$           & 37.0 meV             \\
mass density, $\rho$                                     & 5.31 g/cm$^{3}$      \\
\hline\hline
\end{tabular}
\end{table}

Accordingly, it follows that the comparison of the three values
follows from
%
\begin{equation}
\frac{\tau _{I_{h}}}{\tau_{h}} = \frac{875}{14}
\frac{k_{B}T^{\ast}}{\hbar \omega_{0}} \, \, ,
\end{equation}
%
\begin{equation}
\frac{\tau _{I_{h2}}}{\tau_{h}} = 28 \left(
\frac{k_{B}T^{\ast}}{\hbar \omega_{0}}\right)^{2} \, ,
\end{equation}
%
\begin{equation}
\frac{\tau_{I_{h}}}{\tau_{I}} = \frac{56}{125}
\frac{k_{B}T^{\ast}}{\hbar \omega_{0}} \, \, ,
\end{equation}
and, therefore, their hierarchy depends on the ratio $\hbar \omega
_{0}/k_{B}T^{\ast}$, that is, on the ratio of the energy of
\textsc{lo}-phonon modes to the nonequilibrium thermal kinetic
energy of the carriers. Numerical values of the three Maxwell times
for several values of the carriers' quasitemperature are given in
Table II.

%
\begin{table}[htbp]
\caption{Numerical Values of Maxwell Times for GaAs}
\begin{tabular}{cccccc}
\hline\hline $T^{*}$ (K) & $\theta_{h}$ (ps) & $\theta_{I_{h}}$ (ps)  & $\theta_{I_{h2}}$ (ps) \\
\tableline 100           & 23.7              & 33.76                  & 193                    \\
150                      & 3                 & 6.4                    & 36.5                   \\
200                      & 1.24              & 3.5                    & 20                     \\
300                      & 0.57              & 2.45                   & 19                     \\
500                      & 0.43              & 3.06                   & 17.5                   \\
1000                     & 0.43              & 6.12                   & 3.5                    \\
\hline\hline
\end{tabular}
\end{table}

It can be noticed that these Maxwell times are in order of
picoseconds to femtoseconds. We stress that $T^{\ast}$ is the
carriers' quasitemperature, that is, the nonequilibrium temperature
which is a measure in Kelvin degrees of their nonequilibrium energy
[26].

Inspection of Table II tells us that it is verified that $\theta_{h}
< \theta_{I_{h}} < \theta_{I_{h2}}$, with this ordering being
inverted for values of $\xi$ sufficiently larger than 1, i.e., for
carriers' quasitemperatures sufficiently larger than Einstein
temperature ($T_{E} = \hbar \omega_{0}/k_{B}$) of the \textsc{lo}
phonons, roughly $T^{\ast} > 432$ Kelvin in GaAs.

\newpage
\section{Concluding Remarks}

In the framework of Mesoscopic Hydro-Thermodynamics applied to the
study of the hydrodynamic motion of a classical molecular fluid, and
of the "fluids" of phonons and carriers in intrinsic and doped
semiconductors, we have characterized the so-called Maxwell times,
or, better to say, generalizations of Maxwell original proposal [1]
as described in the Introduction. These Maxwell times are associated
with the fluxes of all orders present in MHT (fluxes of particles --
or quasi-particles -- and of heat). A complete characterization is
obtained in the case of a classical fluid (of molecules, polymers,
etc.), presented in Section II and III. It follows the quite
important result that they satisfy a hierarchy showing that they are
increasingly diminishing as the order of the fluxes increases. This
allows for establishing criteria for the choice of the contraction
of description to be used in each case.

In other words to introduce an appropriate -- for each case --
contraction of description: \emph{this contraction implies in
retaining the information considered as relevant for the problem in
hands, and to disregard irrelevant information} [27].

Elsewhere [24] it has been discussed the question of the contraction
of description (reduction of the dimensions of the nonequilibrium
thermodynamic space of states), where a criterion for justifying the
different levels of contraction is derived: It depends on the range
of wavelengths and frequencies which are relevant for the
characterization, in terms of normal modes, of the
hydro-thermodynamic motion in the nonequilibrium open system. It can
be shown that the truncation criterion \emph{rests on the
characteristics of the hydrodynamic motion that develops under the
given experimental procedure}.

Inclusion of higher and higher-order fluxes implies in describing a
motion involving increasing Knudsen numbers per hydrodynamic mode,
that is, governed by smaller and smaller wavelengths -- larger and
larger wavenumbers -- accompanied by higher and higher frequencies.
In a qualitative manner, we can say that, as a general ``thumb
rule", the criterion indicates that \emph{a more and more restricted
contraction can be used when larger and larger are the prevalent
wavelengths in the motion}. Therefore, in simpler words, when the
motion becomes more and more smooth in space and time the more
reduced can be the dimension of the space of basic macrovariables to
be used for the description of the nonequilibrium thermodynamic
state of the system. It can be conjectured a general criterion for
contraction, namely, a contraction of order $r$ (meaning keeping the
densities and their fluxes up to order $r$) can be introduced once
we can show that in the spectrum of wavelengths, which characterize
the motion, predominate those larger than a ``frontier" one,
$\lambda^{2}_{(r,r+1)} = v^{2}\theta_{r}\theta_{r+1}$ where $v$ is
of the order of the thermal velocity and $\theta_{r}$ and
$\theta_{r+1}$ the corresponding \emph{Maxwell times} associated to
the $r$ and $r+1$ order fluxes.

Section IV has been devoted to the characterization of Maxwell times
in the thermal motion of phonons in intrinsic semiconductors. It is
shown the interesting results that the Maxwell times have an
expression for their inverse in the form of a weighted
Mathiessen-like rule involving the inverse of the relaxation time of
all phonon modes. We can say that it is expected once the
hydrodynamic motion is a collective one composed of the contribution
of all the phonon modes. As in the previous case Maxwell times
follow a hierarchy involving ever decreasing values as the order of
the fluxes increases.

Finally, in Section V, for the case of doped polar semiconductors,
the Maxwell times associated to the hydrodynamic motion of the
electrons in Bloch bands have been obtained. They are mainly
determined by the presence of the polar Fr\"{o}hlich interaction
between electrons and phonons and they follow a hierarchy of values,
which is dependent on the ratio of the phonon energy $\hbar
\omega_{0}$ with the electron thermal (kinetic) energy: it implies
in a series of decreasing values as long as the quasitemperature of
the ``hot" electrons is higher than the Einstein temperature of the
phonons.

\newpage
\textbf{Acknowledgments:} The authors would like to acknowledge
partial financial support received from the S\~{a}o Paulo State
Research Agency (FAPESP) and the Goi\'{a}s State Research Agency
(FAPEG). \vspace{2cm}

\textbf{In Memoriam:} \emph{With very sad feelings, we regret to
report the passing away of our dear colleague \'{A}urea Rosas
Vasconcellos, a genuine, devoted and extremely competent Teacher and
Researcher with fervent dedication to Theoretical Physics in the
Condensed Matter area. She had a very important contribution to this
paper, particularly in what refers to the ``quantum" fluids of
phonons and electrons in semiconductors}.

\newpage


\end{document}